\documentclass[a4paper,11pt]{article}
\pdfoutput=1 

\usepackage{jheppub}
\usepackage[utf8]{inputenc}
\usepackage{pgf}
\usepackage{epsfig}
\usepackage{amsmath}
\usepackage{graphicx}
\usepackage{color}

\def\A0#1{\Pi_{\rm #1}(0)}
\def\AP0#1{\Pi'_{\rm #1}(0)}

\def\be{\begin{equation}}
\def\ee{\end{equation}}
\def\bea{\begin{array}}
\def\eea{\end{array}}
\def\beqa{\begin{eqnarray}}
\def\eeqa{\end{eqnarray}}
\def\beqas{\begin{eqnarray*}}
\def\eeqas{\end{eqnarray*}}

\def\bp{\begin{picture}}
\def\ep{\end{picture}}
\def\bc{\begin{center}}
\def\ec{\end{center}}
\def\bfig{\begin{figure}}
\def\efig{\end{figure}}

\def\bit{\begin{itemize}}
\def\eit{\end{itemize}}
\def\nn{\nonumber}
\def\f{\frac}

\def\[{\left[}
\def\]{\right]}
\def\({\left(}
\def\){\right)}

\def\..{\left.}
\def\.{\right.}
\def\tl{\tilde}
\def\ra{\rightarrow}
\def\la{\leftarrow}

\def\tm{\times}

\def\da{\dagger}

\def\la{\lambda}

\def\al{\alpha}

\def\ka{\kappa}

\def\ep{\epsilon}

\def\pa{\partial}
\def\pr{\prime}

\title{Explaining the Muon g-2 Anomaly in Deflected AMSB for NMSSM}
\author[a]{Li Jun Jia,}
\author[a]{Zhuang Li,}
\author[a]{Fei Wang*\note{*Corresponding author.}}
\affiliation[a]{School of Physics, Zhengzhou University, Zhengzhou 450000, P. R. China}
\emailAdd{feiwang@zzu.edu.cn}
\abstract{We propose to embed the General NMSSM (Next-to-Minimal Supersymmetric Standard Model) into the deflected AMSB (Anomaly Mediated Supersymmetry Breaking) mechanism with Yukawa/gauge deflection contributions. After integrating out the heavy messenger fields, the analytical expressions of the relevant soft SUSY breaking spectrum for General NMSSM at the messenger scale can be calculated. We find that successful EWSB (Electroweak Symmetry Breaking) and realistic low energy NMSSM spectrum can be obtained in some parameter regions. In addition, we find that the muon $g-2$ anomaly and electron $g-2$ anomaly (for positive central value electron $g-2$ experimental data)  can be jointly explained to $1\sigma$ and $2\sigma$ range, respectively. The $Z_3$ invariant NMSSM, which corresponds to $\xi_F=0$ in our case, can also jointly explain the muon and electron anomaly to $1\sigma$ and $2\sigma$ range, respectively.}

\begin{document}
\maketitle \indent
\newpage
\section{Introduction}
The Standard Model (SM) of particle physics is very successful in explaining the vast experimental measurements up to the electroweak (EW) scale, including the discovery of Higgs boson by the Large Hadron Collider (LHC)~\cite{ATLAS:higgs,CMS:higgs}. However, it still has many theoretical and aesthetic problems, for~example, the~quadratic divergence of the fundamental Higgs scalar mass, the~dark matter~(DM) puzzle, and the origin of baryon asymmetry in the universe. In addition, it has been known for a long time that the theoretical prediction of the muon anomalous magnetic moment $a_\mu\equiv (g-2)_\mu/2$ for SM has subtle deviations from the experimental values. In~fact, combining the recent reported E989 muon $g-2$ measurement with the previous BNL result~\cite{g-2:BNL,g-2:PDG}, the~updated world average experimental value of $a_\mu$ is given by~\cite{g-2:FNAL}
\beqa
a^{\rm FNAL+BNL}_\mu = (11659206.2 \pm 4.1) \tm 10^{-10}~,
\eeqa
which has a $4.2\sigma$ deviation from the SM prediction~\cite{g-2:Th}
 \beqa
\Delta a^{\rm{FNAL+BNL}}_\mu =(25.1 \pm  5.9)  \tm 10^{-10}~.
\eeqa

 {In} addition to the reported muon $g-2$ anomaly, the~experimental data on electron $g-2$ also reported some deviations from the SM predictions. From~the measurement of the fine structure constant $\alpha_{\rm em}(\textit{Cs})$ by the Berkeley experiment using $^{133}Cs$ atoms~\cite{Parker:2018vye}, the~experimental value on electron $g-2$ by~\cite{Hanneke:2008tm} has a $2.4\sigma$ deviation from the SM prediction~\cite{Aoyama:2019ryr}
\begin{equation}
        \Delta a_e^{\rm{Exp-SM}} =a_e^{\rm Exp}-a_e^{\rm SM}(\text{$^{133}Cs$})
        =(-8.8\pm 3.6)\times 10^{-13}~,
 \label{eg-2:1}
\end{equation}
with the center value taking the negative sign. Such a result is not consistent with the most accurate 2020 measurement using $^{87}Rb$ atoms~\cite{Morel:2020dww}, which reported a $1.6\sigma$ deviation from SM prediction
\begin{equation}
        \Delta a_e^{\rm{Exp-SM}} =a_e^{\rm Exp}-a_e^{\rm SM}(\text{$^{87}Rb$})
        =(4.8\pm 3.0)\times 10^{-13}~,
\label{eg-2:2}
\end{equation}
with the center value taking the positive sign. Although~the deviation of electron $g-2$ is still controversial, its possible theoretical implications should not be overlooked.  The~previous problems and anomalies strongly indicate that SM should not be the whole story and it only acts as the low energy effective theory of some new physics beyond the~SM.

Various new physics models had been proposed to deal with the problems that bother the SM. Among~them, low-energy supersymmetry~(SUSY) is the most attractive one, which can accommodate almost all the solutions of such problems together in a single framework. In particular, the~discovered 125 GeV Higgs scalar lies miraculously in the {small} {`}$115-135${'} GeV window predicted by the low energy SUSY, which is a strong hint of weak scale SUSY. So, if~low energy SUSY is indeed the new physics beyond the SM, it should account for the new muon $g-2$ anomaly. SUSY explanations of the muon $g-2$ anomaly can be seen in the literatures~\cite{Athron:2021iuf,Du:2022pbp,Li:2021pnt,Crivellin:2021rbq,Endo:2021zal,Gu:2021mjd,VanBeekveld:2021tgn,Yin:2021mls,Abdughani:2021pdc,Cao:2021tuh,Wang:2015nra,Wang:2021bcx,Cox:2018qyi,Yang:2022gvz,Cox:2021nbo,Han:2021ify,Baum:2021qzx,Zhang:2021gun,Ahmed:2021htr,Yang:2021duj,Aboubrahim:2021xfi,Chakraborti:2021bmv,Baer:2021aax,Altmannshofer:2021hfu,Aboubrahim:2021phn,Zhang:2021nzv,Jeong:2021qey,Abdughani:2019wai}. Furthermore, there have already been several discussions offering combined explanations of the experimental results for electron and muon $g-2$ anomaly in the SUSY framework~\cite{emu:3,emu:2,MEG:2016leq,emu:4,Endo:2019bcj,Ali:2021kxa,Yang:2020bmh,Cao:2021lmj,emu:0}. On~the other hand, it is rather non-trivial for the low-energy SUSY breaking spectrum to  be consistent with recent experimental bounds, for~example, the~LHC exclusion bounds and DM null search results. As~the low-energy soft SUSY breaking parameters are fully determined by the predictive SUSY breaking mechanism, their intricate structures can be a  consequence of the mediation mechanism of SUSY breaking in the UV theory, for~example, the~well-motivated anomaly mediated SUSY breaking~(AMSB)~\cite{Randall:1998uk,Giudice:1998xp} mechanism. 

  Minimal AMSB, which is determined solely by the F-term VEV of the compensator superfield  $F_\phi$ (its value is approximately equal to the gravitino mass $m_{3/2}$),
is insensitive to its UV theory~\cite{AMSB:RGE} and predicts a flavor conservation soft SUSY breaking spectrum.
Unfortunately, negative slepton squared masses are predicted and the minimal scenario must be extended. Although~there are many possible ways to tackle such tachyonic slepton masses problems, the~most elegant solution from an aesthetic point of view is the deflected AMSB~(dAMSB)~\cite{Pomarol:1999ie,Rattazzi:1999qg,Okada:2002mv,Okada:2012nr,Wang:2015apa} scenario, which adopts non-trivially an additional messenger sector to deflect the AMSB trajectory and
 push the negative slepton squared masses to positive values by additional gauge mediation~contributions.

 Next-to-Minimal Supersymmetric Standard Model~(NMSSM)~\cite{Ellwanger:2009dp,Maniatis:2009re} can elegantly solve  the $\mu$ problem that bothers the Minimal Supersymmetric Standard Model~(MSSM) with an additional singlet sector. In addition, with~additional tree-level contributions or through doublet-singlet mixing, NMSSM can accommodate easily the discovered 125 GeV Higgs boson mass. However, soft SUSY breaking parameters of low-energy NMSSM from a typical SUSY breaking mechanism, such as gauge-mediated SUSY breaking~(GMSB)~\cite{GMSB,GMSB1}, are always bothered by the requirement to achieve successful EWSB with suppressed trilinear couplings $A_\ka,A_\la$ and $m_S^2$, rendering the model building non-trivial~\cite{NMSSM:GMSB}. Such difficulties always persist in ordinary AMSB-type scenarios.
  We find that the phenomenologically interesting NMSSM spectrum with successful EWSB can be successfully generated by combining both AMSB and GMSB-type contributions. In addition, the~discrepancy between the theoretical predictions for the muon (and electron) anomalous magnetic momentum and the experiments can be explained in such~model.

This paper is organized as follows. In~Section~\ref{section2}, we propose our model and discuss the general expression for soft SUSY parameters. The~soft SUSY parameters for General NMSSM are given in our scenario. The relevant numerical results are studied in Section~\ref{section3}. Section~\ref{section4} contains our conclusions.
\section{Soft SUSY Breaking Parameters of NMSSM from Deflected~AMSB}\label{section2}
We know that NMSSM is well motivated theoretically to solve the $\mu-B\mu$ problem. By~imposing the discrete $Z_3$ symmetry, a~bare $\mu$ term is forbidden in the NMSSM.
An effective $\mu_{eff}$ parameter can be generated by
\beqa
\mu_{eff} \equiv \lambda \left< s \right>,
\eeqa
after the $S$ field acquires the VEV $\left< s \right>$, which lies of order the electroweak scale and breaks the $Z_3$ symmetry.
The $Z_3$ invariant superpotential couplings are given by~\cite{Ellwanger:2009dp,Maniatis:2009re}
\beqa
W_{Z_3 NMSSM}&=&W_{MSSM}|_{\mu=0}+\la S H_u H_d+\f{\ka}{3} S^3~,~
\eeqa
with
\beqa
W_{MSSM}|_{\mu=0}=y^u_{ij} Q_{L,i} H_u U_{L,j}^c- y^d_{ij} Q_{L,i} H_d D_{L,j}^c-y^e_{ij} L_{L,i} H_d E_{L,j}^c~.
\eeqa

   {The} soft SUSY breaking parameters are given as
\beqa
{\cal L}_{\tiny Z_3NMSSM}^{soft}={\cal L}_{MSSM}^{soft}|_{B=0}-\(A_\la \la S H_u H_d+A_\ka \f{\ka}{3} S^3\)-m_S^2 |S|^2~.
\eeqa

  {The} breaking of $Z_3$ discrete symmetry in the scale-invariant NMSSM is bothered with the domain-wall problem, creating unacceptably large anisotropies of the CMB and spoiling successful BBN predictions. In addition, discrete global symmetry cannot be exact at the Planck scale and may be violated by Planck scale suppressed gravitational interactions. Although~one can find complicated solutions to such difficulties, it is interesting to go beyond the $Z_3$ invariant scheme and adopt the general NMSSM~(GNMSSM) to evade the previous problems. In~fact, effective GNMSSM can also be generated from scale invariant NMSSM if one introduces, in~the Jordan frame, a~$\chi H_u H_d$ term in the frame function that couples to the~curvature.

GNMSSM is the most general single gauge singlet chiral superfield $S$ extension of MSSM, including the most general renormalizable couplings in the superpotential and the corresponding soft SUSY breaking terms in ${\cal L}_{soft}$. On~the other hand, as~noted previously, GNMSSM does not adopt the $Z_3$ symmetry, which can contain the $Z_3$ breaking terms~\cite{Ellwanger:2009dp,Maniatis:2009re}
\beqa
W_{\not{Z}_3 NMSSM}&=\xi_F S+\f{1}{2}\mu^\pr S^2+\hat{\mu} H_u H_d~.
\eeqa

 {So}, the~general superpotential of GNMSSM contains
\beqa
W_{GNMSSM}\supseteq W_{Z_3NMSSM}+W_{\not{Z}_3 NMSSM}~,
\eeqa
 with the corresponding new terms of soft SUSY breaking parameter
 \beqa
 -{\cal L}\supseteq m_3^2 H_u H_d+\f{1}{2} m_S^{\pr 2} S^2+\xi_S S+h.c.~.
 \eeqa

 Such low-energy soft SUSY breaking parameters are determined by the corresponding SUSY breaking mechanism in the UV-completion theory~\cite{Ellwanger:2008py,Han:1999jc,GMSB,Nilles:1983ge}. However, it is rather non-trivial to generate the phenomenological desirable low-energy soft SUSY breaking spectrum from UV theory. In~fact, null search results of sparticles  with 139 $fb^{-1}$ of data at the (13 TeV) LHC by the ATLAS and CMS collaborations~\cite{Run2} suggest that the low-energy SUSY spectrum should have an intricate pattern. For~example, the~first two generation squarks need to be heavy to avoid the stringent constraints from LHC. To~explain the muon $g-2$ anomaly, light electroweakinoes and sleptons are always preferred. On~the other hand, the~discovered 125 GeV Higgs boson by both the ATLAS and CMS collaborations of LHC may indicate stop masses of order  5$\sim$10  TeV or TeV scale stops with large trilinear coupling $A_t$. As~light stops can be preferred to keep electroweak naturalness, large trilinear coupling $A_t$ is, therefore, needed to accommodate the 125 GeV Higgs. We know that minimal GMSB predicts vanishing $A_t$ at the messenger scale and introducing messenger--matter interactions will sometimes be bothered with additional flavor constraints. So, we propose to generate the phenomenological desirable low-energy NMSSM soft spectrum in the predictive AMSB-type framework with additional Yukawa mediation~contributions.

As the slepton sector of NMSSM is the same as that of MSSM, the~prediction of the NMSSM soft spectrum from AMSB is still bothered with the tachyonic slepton problem. The~most elegant solution to such a problem is to add an additional messenger sector, which can deflect the AMSB trajectory properly so that the slepton squared masses can be pushed to positive values. The~minimal version of deflected AMSB in NMSSM, which is a straightforward extension of ordinary minimal deflected AMSB for MSSM, adopts no additional interactions other than that between the spurion field and the messengers. However, it is still problematic in realizing successful EWSB and triggering a non-vanishing $\langle s\rangle$, as~the necessary condition $A_\ka^2\gtrsim 9 m_S^2$~\cite{Ellwanger:2009dp,Maniatis:2009re} can not be satisfied easily. In addition, our numerical results indicate that the corresponding SUSY contributions to muon $g-2$ anomaly is very small for the few survived parameter points. So, additional coupling terms in the superpotential involving the singlet $S$ and messengers can be introduced to bring new Yukawa deflection contributions to the soft SUSY breaking spectrum and alter those parameters relevant for~EWSB.

To evade unwanted mixing between the singlet chiral field $S$ and the spurion field $X$, we adopt the  following form of superpotential at the messenger scale
\beqa
W\supseteq  W_{Z_3 NMSSM}+ \sum\limits_{i=1}^{2N}\la_{X;i} X(\bar{\Psi}_i{\Psi}_i)+\sum\limits_{j=1}^{N}\la_{S;j} S(\bar{\Psi}_j{\Psi}_{j+1} )+ W(X)~,
\label{sector:messenger}
\eeqa
with the coupling form first proposed in~\cite{GMSB:NMSSM} for GMSB embedding of~NMSSM.

The soft SUSY breaking parameters in deflected AMSB can be given by mixed anomaly mediation and gauge mediation contributions. When the renormalization group equation~(RGE) evolves down below the messenger thresholds, the~anomaly mediation contributions will receive threshold corrections after integrating
out the messenger fields. The~wavefunction renormalization approach~\cite{wavefunction:hep-ph/9706540,chacko} can also be used in deflected AMSB~\cite{Fei:1602.01699,Fei:1703.10894} to keep track of supersymmetry-breaking effects with
a manifestly supersymmetric formalism and obtain the soft SUSY breaking spectrum. In~deflected AMSB, the~messenger threshold $M_{mess}$ and the RGE scale $\mu$ in the wavefunction superfield and gauge kinetic superfield can be replaced by the following combinations involving the spurious chiral fields $X$
\beqa
 M_{mess}\ra \sqrt{\f{X^\da X}{\phi^\da\phi}}~,~~~~~\mu\ra \f{\mu}{\sqrt{\phi^\da\phi}}~,
\eeqa
which, after~substituting the F-term VEV of $X$ and $\phi$, can obtain the soft SUSY breaking parameters.
Below the messenger scale, the~messenger superfields will be integrated out and will deflected the trajectory from ordinary AMSB trajectory. The~superpotential for pseudo-moduli superfield $W(X)$ can be fairly generic and leads to a deflection parameter of either sign given by
\beqa
d \equiv \frac{F_X}{M F_\phi} -1 .
\eeqa

   {After} integrating out the heavy messengers, we can obtain the soft SUSY breaking spectrum at the messenger scale.
The soft gaugino masses at the messenger scale can be given by
\beqa
M_{i}(M_{mess})&=& g_i^2\(\f{F_\phi}{2}\f{\pa}{\pa \ln\mu}-\f{d F_\phi}{2}\f{\pa}{\pa \ln |X|}\)\f{1}{g_i^2}(\mu,|X|,T)~,
\eeqa
with\vspace{-3pt}
\beqa
\f{\pa}{\pa \ln |X|} g_i(\al; |X|)=\f{\Delta b_i}{16\pi^2} g_i^3~,
\eeqa

   {The} trilinear soft terms can be determined by the wavefunction renormalization factors
\beqa
A_0^{ijk}\equiv \f{A_{ijk}}{y_{ijk}}&=&\sum\limits_{i}\(-\f{F_\phi}{2}\f{\pa}{\pa\ln\mu}+{d F_\phi}\f{\pa}{\pa\ln X}\) \ln \[Z_i(\mu,X,T)\]~,\nn\\
&=&\sum\limits_{i} \(-\f{F_\phi}{2} G_i^- +d F_\phi\f{\Delta G_i}{2}\)~.
\eeqa

   {In} our convention, the~anomalous dimension are expressed in the holomorphic basis~\cite{shih}
\beqa
G^i\equiv \f{d Z_{ij}}{d\ln\mu}\equiv-\f{1}{8\pi^2}\(\f{1}{2}d_{kl}^i\la^*_{ikl}\la_{jmn}Z_{km}^{-1*}Z_{ln}^{-1*}-2c_r^iZ_{ij}g_r^2\).
\eeqa
and $\Delta G\equiv G^+-G^-$ the discontinuity of anomalous dimension across the messenger threshold.  {Here,} {`}$G^+(G^-)${'} denote the values of anomalous dimension above (below) the messenger threshold, respectively. The~soft scalar masses are given by
\begingroup\makeatletter\def\f@size{9.8}\check@mathfonts
\def\maketag@@@#1{\hbox{\m@th\normalsize\normalfont#1}}%
\beqa
m^2_{soft}&=&-\left|-\f{F_\phi}{2}\f{\pa}{\pa\ln\mu}+d F_\phi\f{\pa}{\pa\ln X}\right|^2 \ln \[Z_i(\mu,X,T)\]~,\\
&=&-\(\f{F_\phi^2}{4}\f{\pa^2}{\pa (\ln\mu)^2}+\f{d^2F^2_\phi}{4}\f{\pa}{\pa(\ln |X|)^2}
-\f{d F^2_\phi}{2}\f{\pa^2}{\pa\ln|X|\pa\ln\mu}\) \ln \[Z_i(\mu,X,T)\].\nn\\
\eeqa
\endgroup

  {From} the previous analytic expressions, we can calculate the soft SUSY breaking parameters for MSSM at the messenger scale. The~gaugino masses are calculated to be
\beqa
M_i=F_\phi\f{\al_i(\mu)}{4\pi}\(b_i-d\Delta b_i\)~,
\eeqa
with the corresponding beta function $(b_1~,b_2~,~b_3)=(\f{33}{5},~1,-3)$~\cite{Martin:1997ns} and the changes of beta function for the gauge couplings
\beqa
\Delta(b_1~,b_2~,~b_3)&=&(~2N,~2N,~2N),
\eeqa
with N representing the family of messengers in~(\ref{sector:messenger}).

The trilinear soft terms are calculated to be
\beqa
A_t&=&\f{F_\phi}{16\pi^2}\[\tl{G}_{y_t}\]~,\nn\\
A_b&=&\f{F_\phi}{16\pi^2}\[\tl{G}_{y_b}\]~,\nn\\
A_\tau&=&\f{F_\phi}{16\pi^2}\[\tl{G}_{y_\tau}\]~,\nn\\
A_\la &=&\f{F_\phi}{16\pi^2}\[\tl{G}_{\la}-d \Delta \tl{G}_{\la}\]~,\nn\\
A_\ka &=&\f{F_\phi}{16\pi^2}\[\tl{G}_{\ka}-d \Delta \tl{G}_{\ka}\]~,\nn\\
\f{\xi_S}{\xi_F}&=&\f{1}{2}\f{m_S^{\pr 2}}{\mu^\pr}=\f{1}{3}A_\ka~,\nn\\
\f{m_3^2}{\hat{\mu}}&=&\f{F_\phi}{16\pi^2}\[\tl{G}_{\hat{\mu}}\]~,
\eeqa
with
\beqa
\tl{G}_{\la}&=&4\la^2+2\ka^2+3y_t^2+3y_b^2+y_\tau^2-(3g_2^2+\f{3}{5}g_1^2)~,\nn\\
\tl{G}_{\ka}&=&6\la^2+6\ka^2~,\nn\\
\tl{G}_{y_t}&=&\la^2+6y_t^2+y_b^2-(\f{16}{3}g_3^2+3g_2^2+\f{13}{15}g_1^2)~,\nn\\
\tl{G}_{y_b}&=&\la^2+y_t^2+6y_b^2+y_\tau^2-(\f{16}{3}g_3^2+3g_2^2+\f{7}{15}g_1^2)~,\nn\\
\tl{G}_{y_\tau}&=&\la^2+3y_b^2+4y_\tau^2-(3g_2^2+\f{9}{5}g_1^2)~,\nn\\
\tl{G}_{\hat{\mu}}&=&3y_t^2+3y_b^2+y_\tau^2-(3g_2^2+\f{3}{5}g_1^2)~,
\eeqa
and the discontinuity of the Yukawa beta functions across the messenger threshold
\beqa
\Delta \tl{G}_{\la} &=&5\sum\limits_{j=1}^N\la_{S;j}^2~,\nn\\
\Delta \tl{G}_{\ka}&=&15\sum\limits_{j=1}^N\la_{S;j}^2~.
\eeqa

The sfermion masses can be calculated to be
\beqa
m^2_{{H}_u}~~&=&\f{F_\phi^2}{16\pi^2}\[\f{3}{2}G_2\al^2_2+\f{3}{10}G_1\al^2_1\]
+\f{F_\phi^2}{(16\pi^2)^2}\[\la^2\tl{G}_\la+3y_t^2\tl{G}_{y_t}\]~,~\nn\\
m^2_{{H}_d}~~&=&\f{F_\phi^2}{16\pi^2}\[\f{3}{2}G_2\al^2_2+\f{3}{10}G_1\al^2_1\]
+\f{F_\phi^2}{(16\pi^2)^2}\[\la^2\tl{G}_\la+3y_b^2\tl{G}_{y_b}+y_\tau^2\tl{G}_{y_\tau}\]~,~\nn\\
m^2_{\tl{Q}_{L;1,2}}&=&\f{F_\phi^2}{16\pi^2}\[\f{8}{3} G_3 \al^2_3+\f{3}{2}G_2\al^2_2+\f{1}{30}G_1\al^2_1\]~,~\nn\\
m^2_{\tl{U}^c_{L;1,2}}&=&\f{F_\phi^2}{16\pi^2}\[\f{8}{3} G_3 \al^2_3+\f{8}{15}G_1\al^2_1\]~,~\nn\\
m^2_{\tl{D}^c_{L;1,2}}&=&\f{F_\phi^2}{16\pi^2}\[\f{8}{3} G_3 \al^2_3+\f{2}{15}G_1\al^2_1\]~,~\nn\\
m^2_{\tl{L}_{L;1,2}}&=&\f{F_\phi^2}{16\pi^2}\[\f{3}{2}G_2\al_2^2+\f{3}{10}G_1\al_1^2\]~,~\nn\\
m^2_{\tl{E}_{L;1,2}^c}&=&\f{F_\phi^2}{16\pi^2}\f{6}{5}G_1\al_1^2~,~
\eeqa
with
\begin{gather}
G_i=2N d^2+4N d-b_i~,~\nn\\
(b_1,b_2,b_3)=(\f{33}{5},1,-3)~.
\end{gather}

For the third generation, the~sfermion masses are given by
\beqa
m^2_{\tl{Q}_{L,3}}&=&m^2_{\tl{Q}_{L;1,2}}+F_\phi^2\f{1}{(16\pi^2)^2}\[y_t^2\tl{G}_{y_t}+y_b^2\tl{G}_{y_b}\]~,\nn\\
m^2_{\tl{U}^c_{L,3}}&=&m^2_{\tl{U}^c_{L;1,2}}+F_\phi^2\f{1}{(16\pi^2)^2}\[2y_t^2\tl{G}_{y_t}\]~,\nn\\
m^2_{\tl{D}^c_{L,3}}&=&m^2_{\tl{D}^c_{L;1,2}}+F_\phi^2\f{1}{(16\pi^2)^2}\[2y_b^2\tl{G}_{y_b}\]~,~\nn\\
m^2_{\tl{L}_{L,3}}&=&m^2_{\tl{L}_{L;1,2}}+F_\phi^2\f{1}{(16\pi^2)^2}\[y_\tau^2\tl{G}_{y_\tau}\]~,~\nn\\
m^2_{\tl{E}_{L,3}^c}&=&m^2_{\tl{E}_{L;1,2}^c}+F_\phi^2\f{1}{(16\pi^2)^2}\[2y_\tau^2\tl{G}_{y_\tau}\]~,~
\eeqa
within which we also include the top, bottom and tau Yukawa contributions. Such contributions can not be neglected at large values of $\tan\beta$.

Expression of the soft $m_S^2$ receives both anomaly mediation, Yukawa mediation, and interference contributions
 \beqa
 m_S^2=\Delta_a m^2_{S}+\Delta_Y m^2_{S}~,
 \eeqa
with the pure anomaly mediation part being
\beqa
\Delta_a m^2_{S}&=&\f{F_\phi^2}{(16\pi^2)^2}\[2\la^2\tl{G}_{\la}+2\ka^2\tl{G}_{\ka}\]~,
\eeqa
and the Yukawa deflection part (including the interference terms) being
\beqa
\Delta_Y m^2_{S}&=&\f{ F_\phi^2}{(16\pi^2)^2}\left\{{d^2}\sum\limits_{i=1}^N(5\la^2_{S;i}G_{\la_{S;i}}^+ )-(d^2+2d) \[ 2\la^2 \Delta \tl{G}_\la+2\ka^2 \Delta \tl{G}_\ka\] \right\}~,\nn\\
G_{\la_{S;a}}^+&=&5(\sum\limits_{j=1}^{N}\la_{S;j}^2)+(\la_{X;a}^2+\la_{S;a}^2)+(\la_{X;a+1}^2+\la_{S;a}^2)~.
\eeqa

\section{Joint Explanation of Muon and Electron $g-2$ Anomaly}\label{section3}
\begin{figure}[htb]
\begin{center}
\includegraphics[width=5.4in]{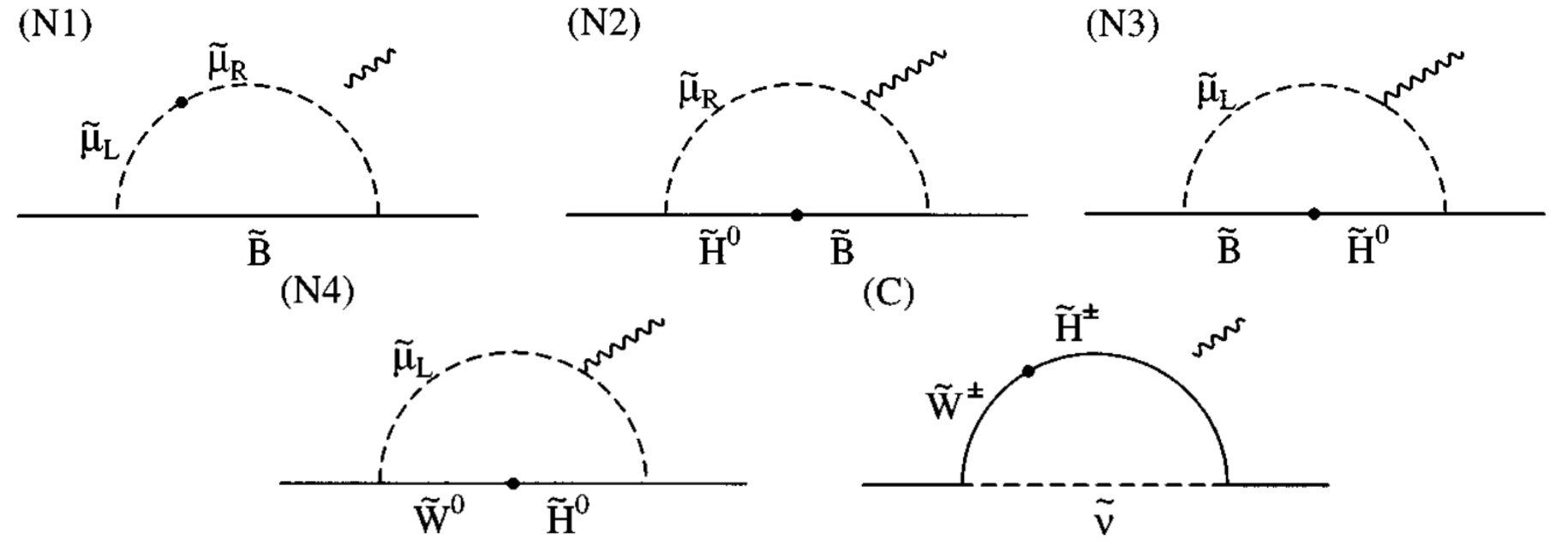}
\vspace{-.5cm}\end{center}
\caption{Leading SUSY contributions to $\Delta a_{\mu}$~\cite{moroi}.}
\label{fig5}
\end{figure}

 We would like to give a joint explanation of muon and electron $g-2$ anomalies in the NMSSM framework from deflected AMSB, taking into account all other low-energy experimental data/exclusion bounds. The~SUSY contributions to muon $g-2$ are dominated by the chargino--sneutrino and the neutralino--smuon loop, and~the corresponding Feynman diagrams are shown in Figure~\ref{fig5}~\cite{moroi}.
At the leading order of $\tan\beta$ and $m_W/m_{SUSY}$, they are evaluated as~\cite{Endo:2013bba}
\begin{align}
  \Delta a_{\mu }(\tilde{\mu }_L, \tilde{\mu }_R,\tilde{B})
 &= \frac{\alpha_Y}{4\pi} \frac{m_{\mu }^2 M_1 \mu}{m_{\tilde{\mu }_L}^2 m_{\tilde{\mu }_R}^2}  \tan \beta\cdot
 f_N \left( \frac{m_{\tilde{\mu }_L}^2}{M_1^2}, \frac{m_{\tilde{\mu }_R}^2}{M_1^2}\right). \label{eq:BmuLR} \\
   \Delta a_{\mu }(\tilde{B}, \tilde{H},  \tilde{\mu }_R)
  &= - \frac{\alpha_Y}{4\pi} \frac{m_{\mu }^2}{M_1 \mu} \tan \beta \cdot
  f_N \left( \frac{M_1 ^2}{m_{\tilde{\mu }_R}^2}, \frac{\mu ^2}{m_{\tilde{\mu }_R}^2} \right), \label{eq:BHmuR} \\
  \Delta a_{\mu }(\tilde{B},\tilde{H},  \tilde{\mu }_L)
  &= \frac{\alpha_Y}{8\pi} \frac{m_\mu^2}{M_1 \mu} \tan\beta\cdot
 f_N
 \left( \frac{M_1 ^2}{m_{\tilde{\mu }_L}^2}, \frac{\mu ^2}{m_{\tilde{\mu }_L}^2} \right),
 \label{eq:BHmuL} \\
 \Delta a_{\mu }(\tilde{W}, \tilde{H},  \tilde{\mu}_L)
 &= - \frac{\alpha_2}{8\pi} \frac{m_\mu^2}{M_2 \mu} \tan\beta\cdot
 f_N
 \left( \frac{M_2 ^2}{m_{\tilde{\mu }_L}^2}, \frac{\mu ^2}{m_{\tilde{\mu }_L}^2} \right),
 \label{eq:WHmuL}  \\
 \Delta a_{\mu }(\tilde{W}, \tilde{H}, \tilde{\nu}_\mu)
 &= \frac{\alpha_2}{4\pi} \frac{m_\mu^2}{M_2 \mu} \tan\beta\cdot
f_C
 \left( \frac{M_2 ^2}{m_{\tilde{\nu }}^2}, \frac{\mu ^2}{m_{\tilde{\nu }}^2}  \right) ,
 \label{eq:WHsnu}
\end{align}

   {Here,} $m_\mu$ is the muon mass, $m_{SUSY}$ the SUSY breaking masses and $\mu$ the Higgsino mass, respectively. The~loop functions are defined as
\begin{align}
&f_C(x,y)= xy
\left[
\frac{5-3(x+y)+xy}{(x-1)^2(y-1)^2}
-\frac{2\log x}{(x-y)(x-1)^3}
+\frac{2\log y}{(x-y)(y-1)^3}
\right]\,,
\\
&f_N(x,y)= xy
\left[
\frac{-3+x+y+xy}{(x-1)^2(y-1)^2}
+\frac{2x\log x}{(x-y)(x-1)^3}
-\frac{2y\log y}{(x-y)(y-1)^3}
\right]\,,
\label{moroi3}
\end{align}
which are monochromatically increasing for $x>0,y>0$ and satisfy $0\le f_{C,N}(x,y) \le 1$. They satisfy $f_C(1,1)=1/2$ and $f_N(1,1)=1/6$ in the limit of degenerate masses. The~SUSY contributions to the muon $g-2$ will be enhanced for small soft SUSY breaking masses and large value of $\tan\beta$. Similar expressions are evident for electron $g-2$ after properly replacing the couplings and mass parameters for muon by those for electrons. The~inclusion of the singlino component in NMSSM will not give sizable contributions to $\Delta a_\mu$ because of the suppressed coupling of singlino to the MSSM sector. However, the~lightest neutral CP-odd Higgs scalar could give non-negligible contributions to $a_\mu$ if it is quite light~\cite{NMSSM:g-2}. The~positive two-loop contribution is numerically more important for a light CP-odd Higgs at approximately 3~GeV and the sum of both one loop and two loop contributions is maximal around $m_{a_1}$$\sim$6~GeV.

It is non-trivial to jointly explain both muon and electron $g-2$ anomalies in a single framework. The~joint explanation of muon and electron $g-2$ anomalies with a negative center value for 
$\Delta a_e$ needs either large non-universal trilinear A-terms~\cite{emu:1,emu:0} or flavor violating off-diagonal elements in the slepton mass matrices~\cite{emu:2,emu:0}. Without~explicit flavor mixings, the~two anomalies can also be explained by arranging the bino--slepton and chargino--sneutrino contributions differently between the electron and muon sectors, requiring heavy left-hand smuon~\cite{emu:3} or light selectrons, wino, and heavy higgsino~\cite{emu:4}. Without~large flavor violation in the lepton sector, we anticipate that the new physics contributions to the leptonic $g-2$ will, in general, scale with the corresponding lepton square masses. Although~such scaling solutions cannot  explain the electron $g-2$ data in~(\ref{eg-2:1}) with negative central value to $2\sigma$ range, it, however, can be consistent with the most accurate data in~(\ref{eg-2:2}) with a positive central value. In~our model, universal soft SUSY breaking parameters are predicted for the muon and the electron sector at the messenger scale. Two loop RGE running will only split slightly the low energy spectrum for the two sectors at the SUSY scale.  Therefore, we anticipate that the scaling solution will approximately hold in our~case.

We use NMSSMTools 5.6.2~\cite{Allanach:2008qq,Ellwanger:2008py} to scan the whole parameter space to find the desired parameter regions that can account for the muon and electron $g-2$ anomaly. In~our numerical calculations, the~choices $N=1$ and $\la_{X;a}=\la_X,\la_{S;a}=\la_S$ are adopted for simply. In addition, although~there are three new free parameters in the GNMSSM superpotential in comparison to $Z_3$ invariant NMSSM, we adopt the most predictive choice with only vanishing $\xi_F$ and other parameters $\hat{\mu}=\mu^\pr=0$. Non-vanishing $\xi_F$ will lead to additional tadpole terms for $S$ and alter the values of $\langle s\rangle$, $\langle h_u\rangle$, and 
$\langle h_d\rangle$ for the EWSB~minima.

The ranges of the input parameters at the messenger scale $M_{mess}$ are chosen to satisfy
 \beqa
&&~ 10^{5} {\rm GeV}< M_{mess}< 10^{15} {\rm GeV},~30 {\rm TeV} <F_\phi< 500 {\rm TeV},~-5<d<5,~0\leq \xi_F\leq F_\phi~,\nn\\
&&~~~~~~0<|\ka|,\la<0.7~~~~{\rm  with}~~\la^2+\ka^2\lesssim 0.7,~~0<\la_X,\la_S<\sqrt{4\pi},~\nn
\label{input:A}
 \eeqa
with $F_\phi\ll M_{mess}$. Below~the messenger scale, the~heavy messenger fields are integrated out and the corresponding RGE trajectory reduces to ordinary GNMSSM type, which can be seen to be deflected from the ordinary AMSB trajectory~\cite{Randall:1998uk,Pomarol:1999ie} and can, possibly, push the tachyonic slepton masses to positive values. So, the~renormalization group equations of GNMSSM~\cite{Ellwanger:2008py} are used to evolve the soft SUSY parameters from the messenger scale to the EW~scale.

 We also impose the following constraints from low energy experimental data other than that already encoded in the NMSSMTools package
  \bit
 \item[(i)] The CP-even component $S_2$ in the Goldstone{-}{`}$eaten${'} combination of $H_u$ and $H_d$ doublets corresponds to the SM Higgs.  Such an $S_2$ dominated CP-even scalar should lie in the combined mass range for the Higgs boson, $122 {\rm GeV}<M_h <128 {\rm GeV}$~\cite{ATLAS:higgs,CMS:higgs}. Note that the uncertainty is 3 GeV instead of default 2 GeV because large $\lambda$ may induce additional ${\cal O}(1)$ GeV correction to $M_h$ at the  two-loop level~\cite{NMSSM:higgs2loop}.

  \item[(ii)] Direct search bounds for low mass and high mass resonances at LEP, Tevatron, and LHC by using the package HiggsBounds-5.5.0~\cite{higgsbounds511}.
  \item[(iii)] Constraints on gluino and squark masses from the latest LHC data~\cite{Heisterkamp:2012qea,ATLAS:2017kyf,CMS:2017arv,ATLAS:2017vjw}
  and the lower mass bounds of charginos and sleptons from the LEP~\cite{LEPmass} results.
  \item[(iv)] Constraints from B physics, such as $B \to X_s \gamma$, $B_s \to \mu^+ \mu^-$and $B^+ \to \tau^+ \nu_\tau$, etc.~\cite{BaBar:2012fqh,BaBar:2012obs,LHCb-BsMuMu,Btaunu}
\begin{eqnarray}
        3.15\times10^{-4} <&Br(B_s\to X_s \gamma)&< 3.71\times10^{-4}~, \\
         1.7\times10^{-9} <&Br(B_s\to\mu^+\mu^-)&< 4.5\times10^{-9}~,  \\
        0.78\times10^{-4} <&Br(B^+\to\tau^+\nu_\tau)&< 1.44\times10^{-4}~.
      \end{eqnarray}
  \item[(v)]  Vacuum stability constraints on the soft SUSY breaking parameters adopted in~\cite{Du:2022pbp}, including the semi-analytic bounds for non-existence of a deeper charge/color breaking (CCB) minimum~\cite{Kitahara:2013lfa} and/or a meta-stable EW vacuum with a tunneling lifetime longer than the age of the universe~\cite{vs2}.

  A sufficient condition to ensure vacuum stability at the EW scale is the requirement that EW vacuum is the global minimum (true vacuum) of the scalar potential. If~the EW vacuum is a local minimum (false vacuum), the~relevant parameter regions can still be allowed if the false EW vacuum is meta-stable with a lifetime longer than the age of the universe.
\item [(vi)] The relic density of the dark matter should satisfy the Planck result $\Omega_{DM} h^2= 0.1199\pm 0.0027$ \cite{Planck} in combination with the WMAP data~\cite{WMAP} (with a $10\%$ theoretical uncertainty).
\eit

\begin{figure}[htb]
\begin{center}
\includegraphics[width=2.7in]{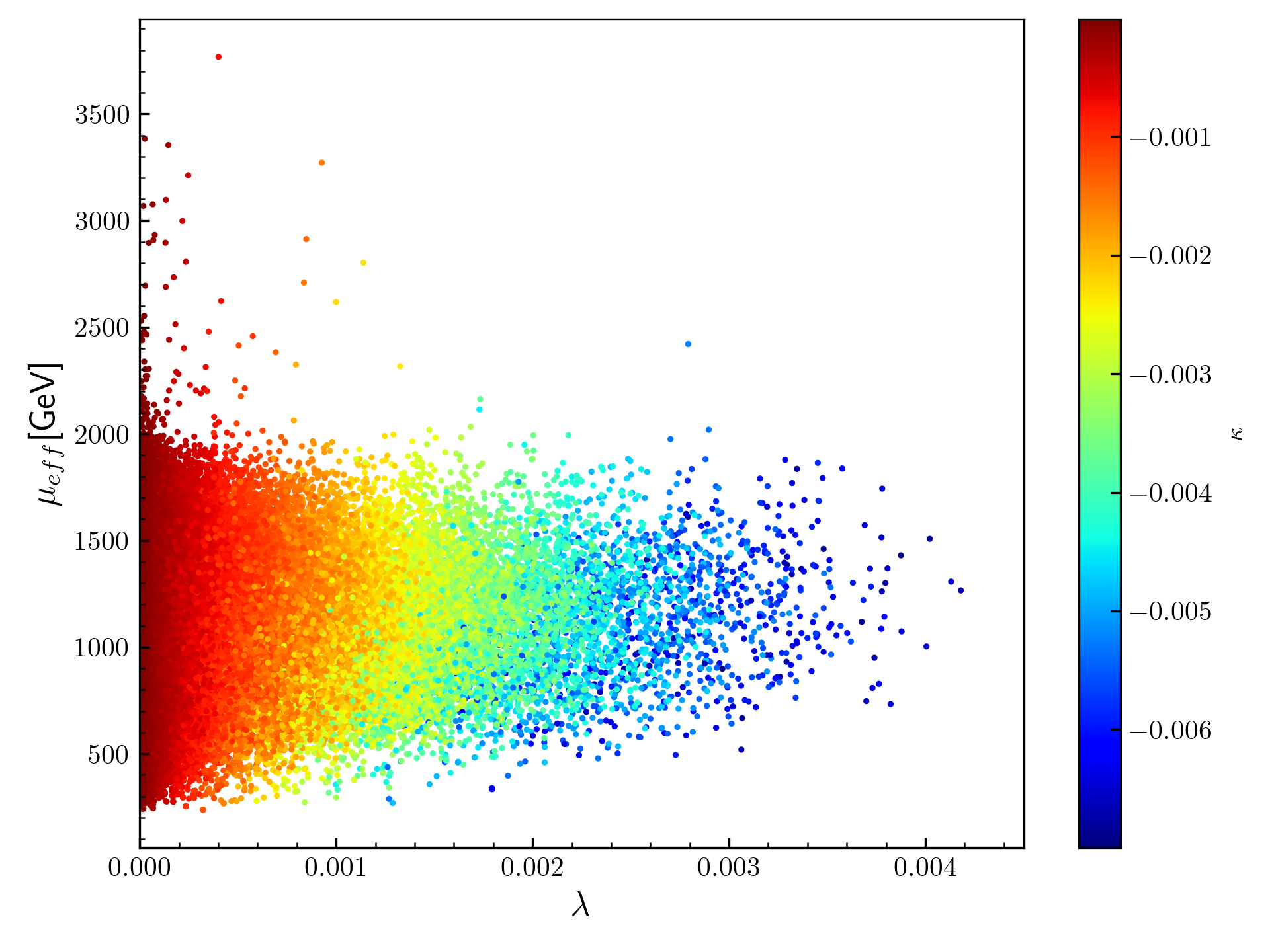}
\includegraphics[width=2.7in]{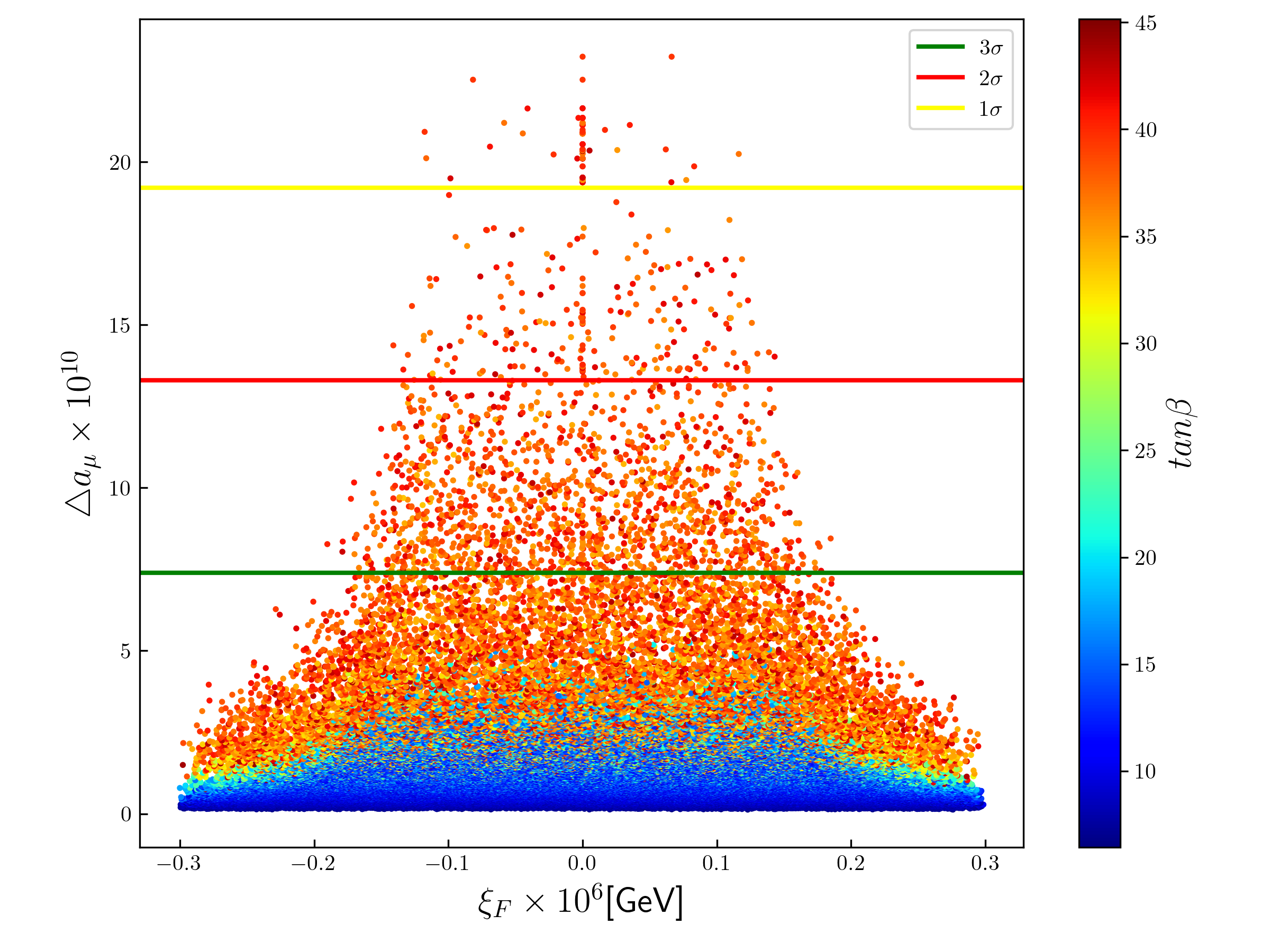}\\
\vspace{-.5cm}\end{center}
\caption{ Survived points that can satisfy the constraints (i--vi). The~allowed ranges of $\lambda$ versus $\mu_{eff},\kappa$ are shown in the left panel while the values of $\xi_F$ versus $\Delta a_\mu,\tan\beta$ are shown in the right~panel.}
\label{fig1}
\end{figure}

We have the following discussions on our numerical results.
\bit
\item Although it is fairly non-trivial to realize successful EWSB in NMSSM from predictive UV-completion models, for~example in ordinary GMSB, numerical scan indicates that some parameter points can still survive the EWSB conditions in our case.  In~fact, additional couplings in the superpotential involving the singlet $S$ and messengers can change the AMSB predictions of $m_S^2$ and $A_\la,A_\ka$ so as that the necessary condition $A_\ka^2\gtrsim 9 m_S^2$ for $\langle s\rangle\neq 0$ and other EWSB conditions can be satisfied. The~values of $\tan\beta$ at the EW scale can be obtained iteratively after we minimize the scalar potential to obtain $\langle s\rangle$. The~allowed values of $\la,\ka$ and the corresponding $\mu_{eff}$ are shown in the left panel of Figure~\ref{fig1}. We can see that the allowed values of $\lambda$ and $\kappa$ are always not large. The~dependence of $\xi_F$ versus the low scale $\tan\beta$ are also shown in the right panel of Figure~\ref{fig1}. An~interesting observation is that successful EWSB can still be allowed with $\xi_F=0$, which is just the $Z_3$-invariant NMSSM case.

\begin{figure}[htb]
\begin{center}
\includegraphics[width=2.7in]{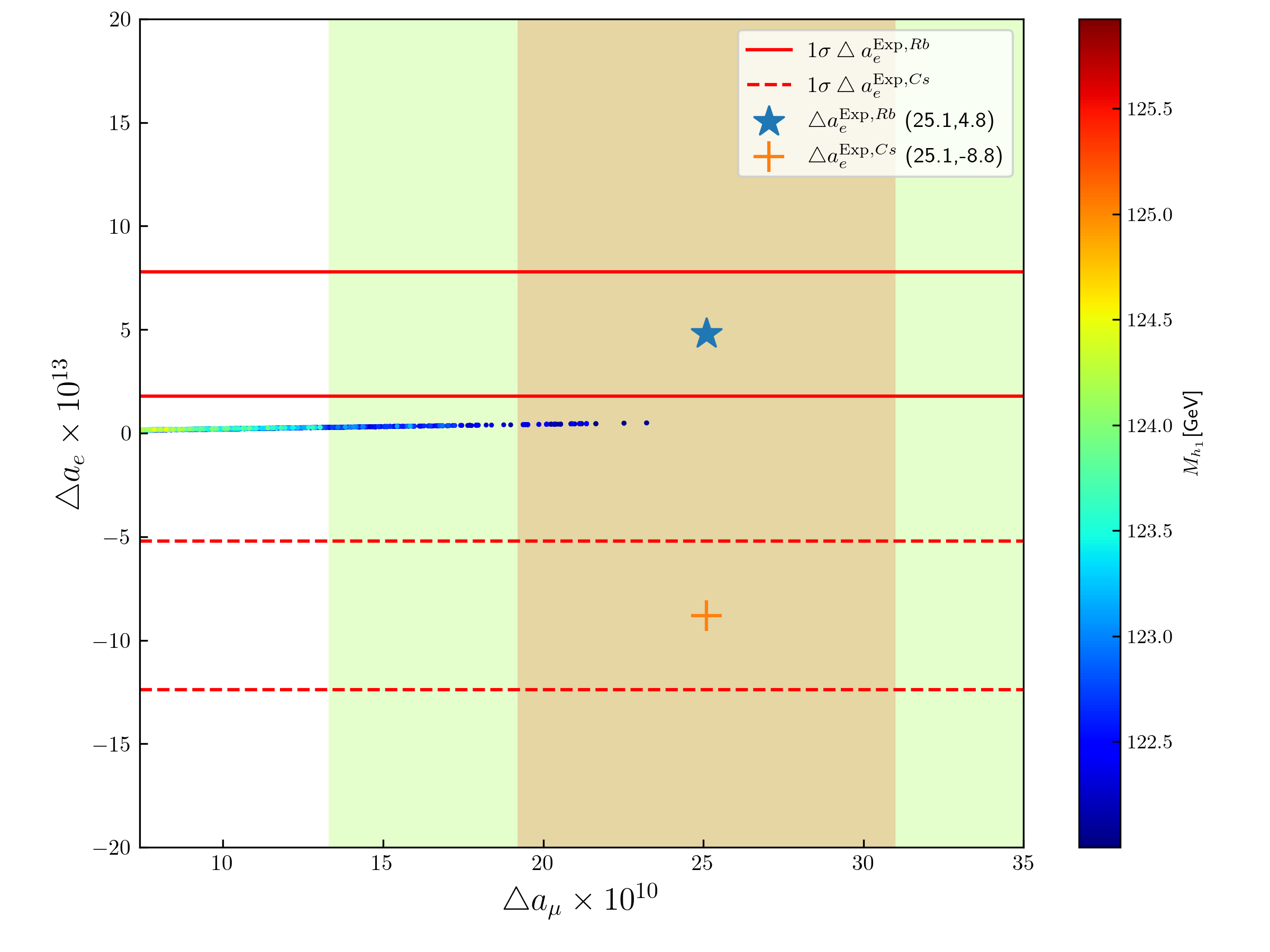}
\includegraphics[width=2.7in]{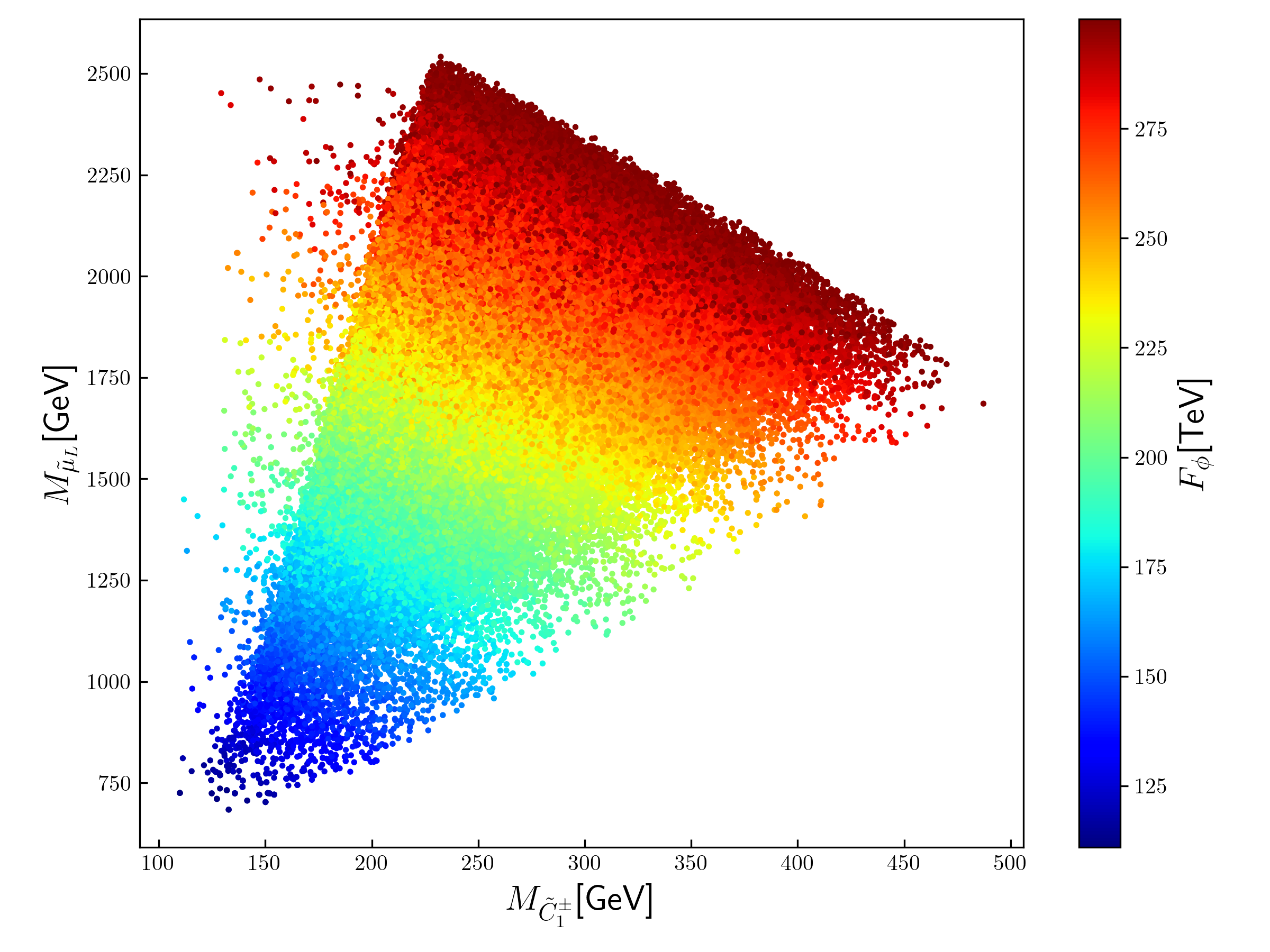}\\
\vspace{-.5cm}\end{center}
\caption{  {In} the left panel, we plot the SUSY contributions to $\Delta a_\mu$ versus $\Delta a_e$ in our case. The~red, green, and white areas represent the $1\sigma,~2\sigma,$ and $3\sigma$ ranges of $\Delta a_\mu$, respectively. The~positive center value case and negative center value case of electron $g-2$ experimental data are shown {with}  {`}$\bigstar${'} {and} {`}$+${'}, respectively. The~slepton masses versus the chargino masses are shown in the right~panel.  }
\label{fig3}  
\end{figure}

\item From our numerical results, we can see in the right panel of Figure~\ref{fig3} that the muon $g-2$ anomaly can be explained to $1\sigma$ range. As~noted previously, small flavor violation in the lepton sector will predict that $\Delta a_e$ and $\Delta a_\mu$ satisfy the scaling relation
\beqa
 \f{\Delta a_e}{\Delta a_\mu}\sim \f{m_e^2}{m_\mu^2}\sim 2.4\tm 10^{-5}~,
\label{eg:1}
 \eeqa
which predicts the same sign of $\Delta a_e$ as that of $\Delta a_\mu$. An explanation of the muon $g-2$ anomaly can also lead to the explanation of the electron $g-2$ anomaly in $2\sigma$ range for positive central value electron $g-2$ experimental data in (\ref{eg-2:1}). Figure~\ref{fig3} shows a scatter plot of $\Delta a_e$ and $\Delta a_\mu$ with the corresponding SM-like Higgs masses in different colors, as~the SM-like Higgs mass always exclude a large portion of otherwise allowed parameter regions. From~(\ref{eg:1}), due to their dependences on the square of the corresponding lepton masses, $\Delta a_e$ can be seen to be of order $10^{-14}$ when $\Delta a_\mu \sim \mathcal{O}(10^{-9})$.
However, $\Delta a_e \sim \mathcal{O}(10^{-14})$ leads to an apparent horizontal line, when the plot for $\Delta a_e$ versus $\Delta a_\mu$ is shown in Figure~\ref{fig3}.

 The NMSSM specific contributions to $\Delta a_{\mu,e}$ are dominantly given by the Barr--Zee type two-loop contributions involving the lightest CP-odd scalar ${a}_1$. However, our numerical results indicate that the $a_1$ relevant NMSSM specific contributions to $\Delta a_{\mu,e}$ are always small and~subdominant.

    The plot of the SUSY contributions to muon anomalous magnetic momentum $\Delta a_\mu$ versus the $\xi_F$ parameter are shown in the right panel of Figure~\ref{fig1}. It can also be seen from the panel that the $Z_3$-invariant NMSSM case, which corresponds to $\xi_F=0$, can also explain the muon $g-2$ anomaly to $1\sigma$ range (and the electron $g-2$ anomaly to $2\sigma$ range by scaling relations).

\begin{figure}[htb]
\begin{center}
\includegraphics[width=2.7in]{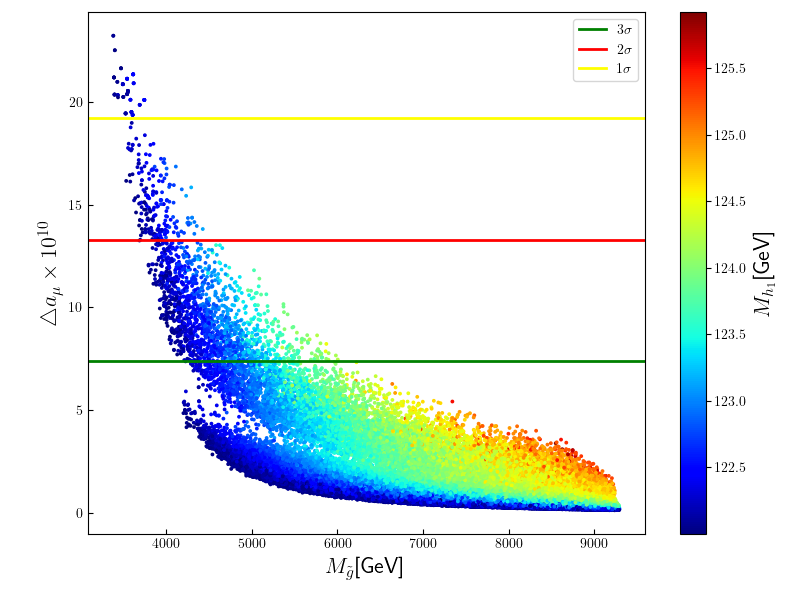}
\includegraphics[width=2.7in]{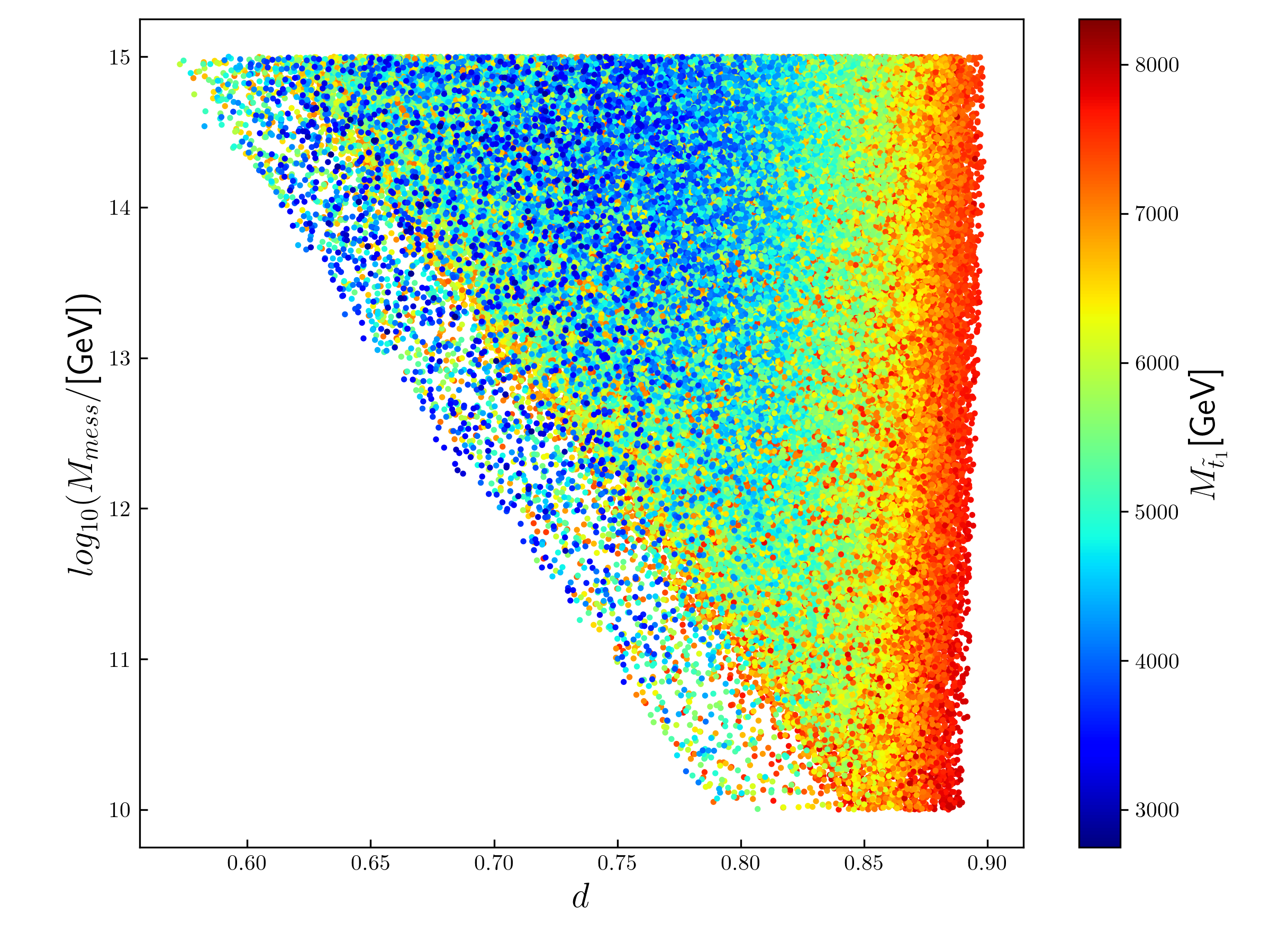}\\
\vspace{-.5cm}\end{center}
\caption{ The SUSY contributions to the muon anomalous magnetic momentum $\Delta a_\mu$ versus the gluino mass $M_{\tl{g}}$ (and the SM-like Higgs mass $M_{h_1}$) are shown in the left panel. The~deflection parameter $d$ versus the messenger scale $M_{mess}$, etc., are shown in the right~panel. }
\label{fig2}
\end{figure}

\item The left panel of Figure~\ref{fig2} shows the plot of the SUSY contributions to muon anomalous magnetic momentum $\Delta a_\mu$ versus the gluino mass $M_{\tl{g}}$. In~AMSB-type scenarios, the~$F_\phi$ parameter determines the mass scales of all the soft SUSY breaking parameters. The~larger the value of $F_\phi$, the~heavier the sfermion and the gaugino masses. We know that light sleptons and electroweakinoes with masses below 0.5$\sim$1  TeV are preferred to explain the muon/electron $g-2$ anomaly via chargino--sneutrino and the neutralino--smuon loops. So, the~SUSY explanations of muon/electron $g-2$ anomalies prefer smaller $F_\phi$, consequently imposing an upper bounds on the low-energy sparticle masses. Our numerical results show that the gluino masses are bounded to lie  $3.5~{\rm TeV}\leq M_{\tl{g}}\leq 6.0~{\rm TeV}$ if the muon $g-2$ anomaly is explained upon $3\sigma$ range. Gluino masses upon 3.5 TeV can possibly be discovered in the future 100 TeV FCC-hh~collider.

    In mSUGRA type models with universal gaugino masses at the GUT scale or GMSB type models, the~gaugino ratios at the EW scale are always given by \mbox{$M_1$:$M_2$:$M_3$ $\approx$ 1:2:6.} Given the LHC exclusion bound  2.2 TeV for $M_{\tl{g}}$ by LHC, such gaugino ratios are not consistent with very light electroweakinoes, making the explanations of the muon $g-2$ anomaly rather hard. In~our case, the~gaugino mass ratios change approximately to
     $M_1$:$M_2$:$M_3$ $\approx$ $\(6.6-2 d\)$:$2\(1-2 d\)$:$6 \(-3-2 d\)$  at the EW scale. Therefore, with~a proper range of deflection parameter $d$, the~gluino mass can be heavy without contradicting the requirements of light electroweakinoes by the explanation of the muon $g-2$ anomaly. We should note that a positive deflection parameter $d$ is always favored to solve the tachyonic slepton problem for few messenger species in deflected AMSB. To~tune the slepton squared masses to small positive values, the~range of $d$ are constrained to lie in a small range. In~fact, our numerical results indicate that the deflection parameters, which parameterizes the relative size between the anomaly mediation contributions and the gauge/Yukawa mediation contributions, are constrained to lie $0.55<d<0.9$ (see the right panel of Figure~\ref{fig2}), allowing the gluino to be heavier than 4 TeV for ${\cal O}(100)$ GeV~wino.

\item It can be seen from the previous figures that the observed SM-like 125 GeV Higgs can be accommodated easily in our model. Additional tree-level contributions to SM-like Higgs mass from NMSSM in general allow much lighter stop masses in comparison to MSSM. In addition, the~trilinear coupling $A_t$ are always predicted to be large in deflected AMSB-type models, which are welcome to give sizeable contributions to the SM-like Higgs mass. Light stops and large $A_t$ can also improve the naturalness measurements of the theory. On~the other hand, the~positive value of $A_t$ tends to decrease to zero and further to large negative values when it RGE evolves down from high input scale to EW scale~\cite{1112.3068}. So, the~values of $A_t$ at the EW scale may not be large for a mildly large messenger scale $M_{mess}$, making the $A_t$ contribution to the SM-like Higgs mass not important for some range of $M_{mess}$. Therefore, the~stop masses are always not light because the allowed values of $\lambda$ are small, leading to small tree-level contributions to the SM-like Higgs mass. It can be seen from Figure~\ref{fig2} that the Higgs mass can be as high as $123.7$ GeV ($124.5$ GeV) and the messenger scale $M_{mess}$ is constrained to be larger than $3\tm 10^{11}$ GeV ( $1.2\tm 10^{10}$ GeV) if the muon $g-2$ anomaly is explained upon $2\sigma$ ($3\sigma$) level, respectively.

    As a comparison, the~Higgs mass is upper bounded to be 118 GeV (120 GeV) when the muon $g-2$ anomaly is explained at $2\sigma$ ($3\sigma)$ level in the CMSSM/mSUGRA, because~light sleptons also indicate light stops (with an universal $m_0$ input at GUT scale), leading to small loop contributions to Higgs masses. So, our deflected AMSB realization of NMSSM is much better in solving the muon $g-2$ anomaly than that of minimal gravity mediation realization of~MSSM.

\begin{figure}[htb]
\begin{center}
\includegraphics[width=2.7in]{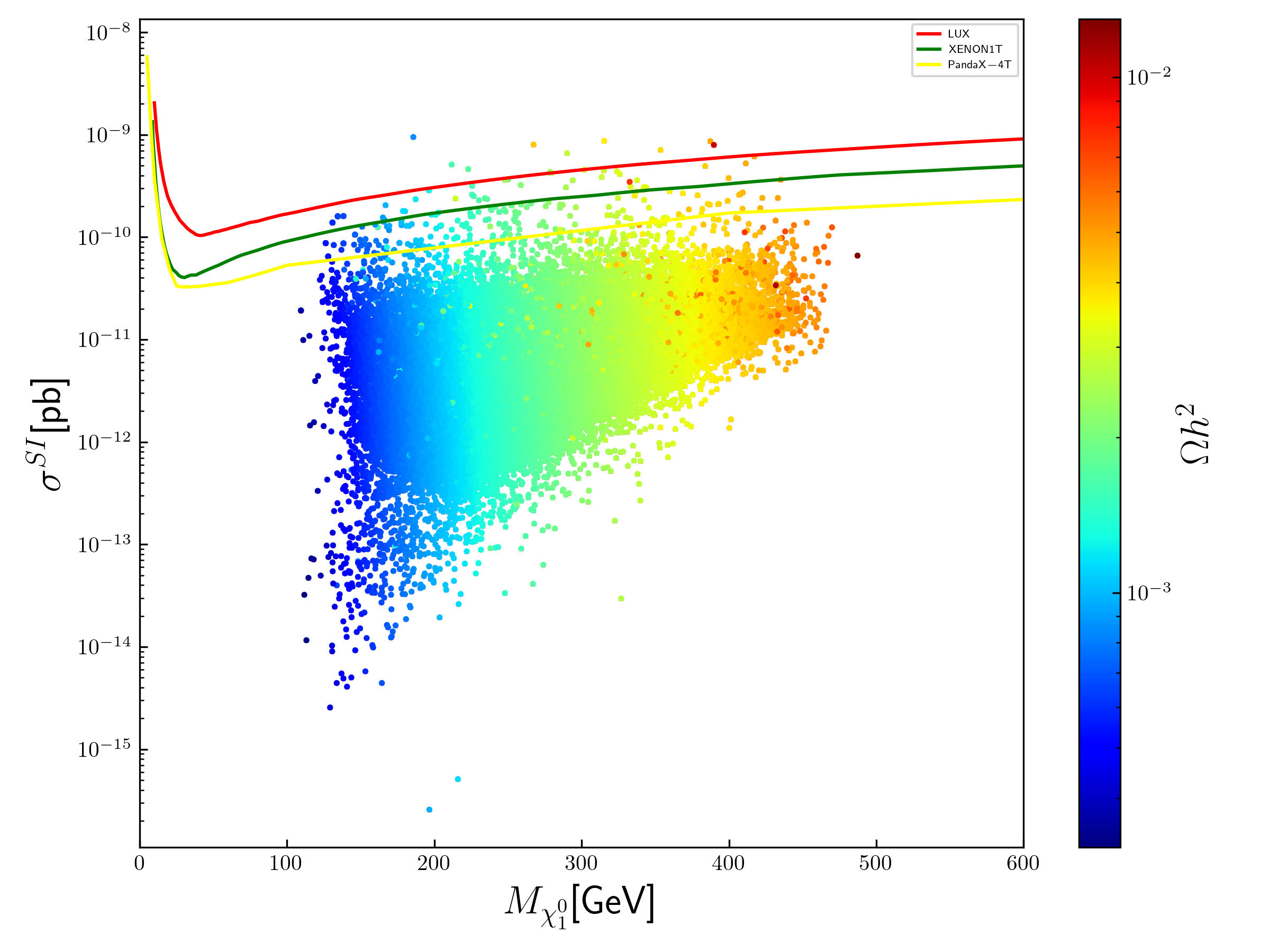}
\includegraphics[width=2.7in]{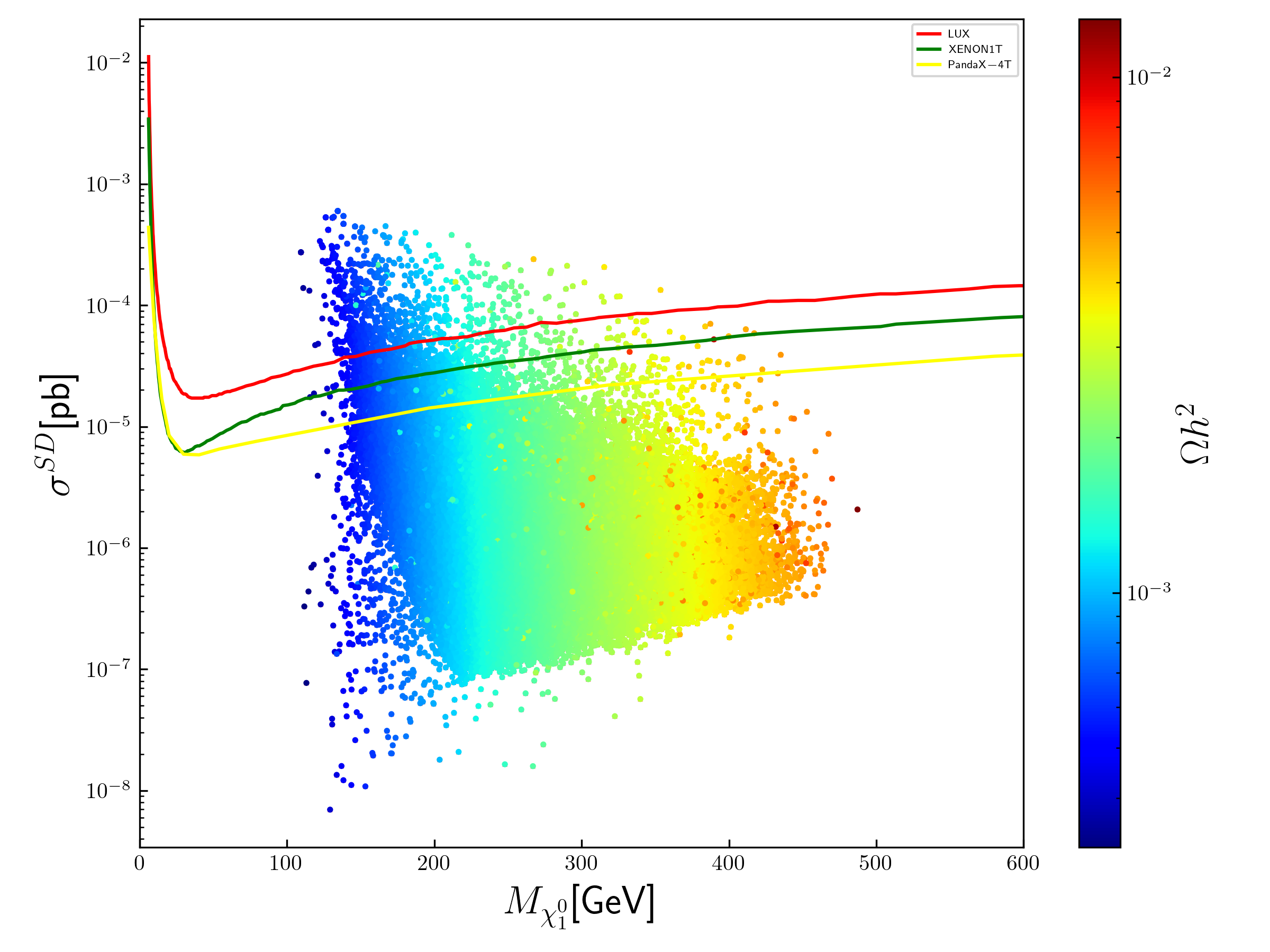}\\
\vspace{-.5cm}\end{center}
\caption{ The Spin-Independent~(SI) (left panel) and Spin-Dependent~(SD) (right panel) DM direct detection bounds for the survived points. The~corresponding DM relic density is shown with different~colors. }
\label{fig4}
\end{figure}

\item Our numerical results indicate that the lightest neutralino DM is always wino-like, which can annihilate very efficiently and lead to the under abundance of DM unless the DM particle mass is heavier than 3 TeV. The~NMSSM-specific singlino component is negligibly small, which, therefore, will not play an important role in DM annihilation processes. Our numerical results indicate that the DM particle is constrained to be lighter than 500 GeV. Therefore, additional DM components, such as the axino, are always needed to provide enough cosmic DM. We also check (see the figures in Figure~\ref{fig4}) that the Spin-Independent~(SI) and Spin-Dependent~(SD) DM direct detection constraints, for~example, the~LUX~\cite{LUX:2016ggv}, XENON1T~\cite{XENON:2018voc,XENON:2019rxp}, and PandaX-4T~\cite{PandaX-4T:2021bab,PandaX:2022xas}, can be satisfied for a large portion of survived points.
\eit
\unskip

\section{Conclusions}\label{section4}
Realistic deflected AMSB-type models are always predictive and can easily lead to light slepton masses when the tachyonic slepton problems that bother them are solved. In addition, the~predicted gaugino mass ratios at the EW scale in deflected AMSB are given by $M_1$:$M_2$:$M_3$ $\approx$ $\(6.6-2 d\)$:$2\(1-2 d\)$:$6 \(-3-2 d\)$, which are different to the ratios \linebreak  \mbox{$M_1$:$M_2$:$M_3$ $\approx$ $1$:$2$:$6$} that appeared in mSUGRA/CMSSM and GMSB. Therefore, much lighter electroweakinoes can still be consistent with the LHC 2.2 TeV gluino lower mass bound. Therefore, the~soft SUSY breaking spectrum from deflected AMSB type scenarios with light sleptons and light electroweakinoes are always favored to solve the muon and electron $g-2$ anomalies. As~the AMSB type scenarios always predict insufficient wino-like DM, additional DM species, such as the NMSSM-specific singlino component, can be welcome to provide additional contributions to the DM relic density. Therefore, the~embedding of NMSSM framework into realistic AMSB-type UV theory is fairly interesting and~well-motivated.

In this paper, we propose to provide a joint explanation of electron and muon $g-2$ anomalies in UV SUSY models in the framework of the anomaly mediation of SUSY breaking. We embed the General NMSSM into the deflected AMSB mechanism with Yukawa/gauge deflection contributions and obtain the relevant soft SUSY breaking spectrum for General NMSSM. After integrating out the heavy messenger fields, the~analytical expressions of the relevant soft SUSY breaking spectrum for General NMSSM at the messenger scale can be calculated, which can be RGE evolved to EW scale with GNMSSM RGE equations. Our numerical scan indicates that successful EWSB and realistic low-energy NMSSM spectrum can be obtained in some parameter regions. Furthermore, we find that, adopting the positively central value electron $g-2$ experimental data, it is possible to jointly explain the muon $g-2$ anomaly and the electron $g-2$ anomaly within a range of 1$\sigma$ and 2$\sigma$, respectively. The~$Z_3$ invariant NMSSM, which corresponds to $\xi_F=0$ in our case, can also jointly explain the muon and electron anomaly to $1\sigma$ and $2\sigma$ range, respectively.

\begin{acknowledgments}
 This work was supported by the
Natural Science Foundation of China under grant numbers 12075213; by the Key Research Project of Henan Education Department for colleges and universities under grant number 21A140025.
\end{acknowledgments}

\end{document}